# Kinetics of phase transformations with heterogeneous correlated-nucleation

By

Massimo Tomellini and Sara Politi

*Dipartimento di Scienze e Tecnologie Chimiche Università di Roma Tor Vergata*

*Via della Ricerca Scientifica 00133 Roma, Italy*

## Abstract

We develop a stochastic approach for describing 3D-phase transformations ruled by time-dependent correlated nucleation at solid surfaces. The kinetics is expressed as a series of correlation functions and, at odds with modeling based on Poisson statistics, it is formulated in terms of actual nucleation rate. It is shown that truncation of the series up to second order terms in correlation functions provides a very good approximation of the kinetics. The time evolution of both total amount of growing phase and surface coverage by the new phase have been determined. The theory is applied to describe progressive nucleation with parabolic growth under time dependent hard-disk correlation. This approach is particularly suitable for describing electrochemical deposition by nucleation and growth where correlation effects are significant. In this ambit the effect of correlated nucleation on the behavior of kinetic quantities used to study electrodeposition has also been investigated.

## 1-Introduction

Theoretical modeling of phase transformations taking place by nucleation and growth with correlated nuclei has in recent years attracted considerable interest from researchers [1-8]. This nucleation process occurs in diffusional growth where diffusion field gives rise to reduced nucleation probability around growing nuclei [9,10]. Correlated nucleation is also induced in stress-driven transformations where the strain field imparts a certain degree of spatial order to the nuclei [11,12]. Until now, the problem has been tackled in the case of homogeneous and isotropic phase transformations in 2D and 3D space, namely in systems that are translationally invariant and isotropic in the whole space where the transformation takes place. These models share the same hypotheses with the celebrated Kolmogorov-Johnson-Mehl-Avrami (KJMA) theory [13-15], with an exception to the nucleation process which is not random throughout the untransformed phase. The KJMA theory, originally developed for first order phase transitions, has been applied to several scientific fields which range from Materials Science to Electrochemistry [16-22] and from Biology to Pharmaceutics [23-25], just to cite a few examples.

The simplest nucleation process is called site saturation (or simultaneous) where all nuclei start growing at the same time [26]. The case of simultaneous nucleation has been firstly modeled in refs.[2-3] for 2D transformations with hard-disk correlation among nuclei. The analytical solution has been found to be in good agreement with computer simulations on 2D lattice [2,3]. Also, computer simulations of non-random distribution of nuclei in either a periodic or a cluster arrangement have been performed for simultaneous nucleation in ref.[7]. Recently, the problem has been studied in more general way in ref.[8] for spatially-decaying correlation and site saturation nucleation. The more involved case of non-simultaneous nucleation has been tackled in refs.[4,5,6] for both time-independent and time-dependent hard-disk correlations. In order to study the stochastic process of dots linked to nucleation, the modelings above quoted make use of correlation-functions. In addition, worthy to note is the alternative approach proposed in ref.[4] which exploits a differential method, with properly defined extended quantities, recovering Avrami's kinetics in the random case [14].

The possibility to deal with spatial correlation is also significant in connection with the growth law of nuclei. On one hand, the KJMA theory does not apply to parabolic-type growths because of the overgrowth of phantom nuclei [14,27]. On the other hand, the possibility to describe correlated nucleation allows one to formulate the theory in terms of *actual* nuclei that are spatially correlated. As a consequence, the restriction on the growth law above mentioned is relaxed since phantom nuclei are eliminated from the formulation of the theory [28-30].

Besides the growth in homogeneous systems it is also of importance the transformation in 3D systems where nucleation is constrained on a surface. For instance, such a process is encountered in Materials Science in the ambit of nucleation and growth at grain-boundaries and interfaces [11,31,32]. It also takes place in film growth by condensation of gas phase at solid surfaces, as studied to a certain extent in the literature [33]. Another technique widely employed to grow films on substrates is based on electrochemical nucleation and growth, where deposition takes place from the liquid phase. In this case the growth is ruled by the diffusion of species in the solution. On the experimental side, the kinetics of deposition is studied by recording chronoamperometric curves which are linked to the amount of deposited species through Faraday's law [9, 22]. It follows that in these systems the modeling of the current-time behavior requires the determination of the kinetics of film growth in terms of total *amount* of deposited material. Until now this modeling has been limited to the case of nucleation which is compliant with the KJMA approach [34]. However, as anticipated above, correlation effects among nuclei are significant in electrochemical nucleation. Due to diffusional growth by mass transport in solution, *actual nuclei* are spatially correlated to an extent that is greater than that occurring in KJMA-type transformations. Experimental data on spatial distribution of actual nuclei in electrochemical deposition are well described by the exclusion zone model for nucleation [9,35-38], namely by considering a region around each nucleus, with size greater than the nucleus size (projected on surface), where further nucleation is prevented. It is also in this field that the present work finds application.

The kinetic theory for phase transformations with time dependent non-Poissonian heterogeneous nucleation has not been proposed so far. The purpose of the present work is to bridge the gap between theoretical modeling of phase transformations ruled by heterogeneous nucleation with random and correlated nuclei. With respect to the random case the present theory has also the advantage to be formulated in terms of actual nucleation, a quantity which is experimentally accessible.

The paper is organized as follows. The first section is devoted to correlation functions and probability theory. In the second one the theory is employed to model the kinetics of 3D diffusional growth of hemispheres. The third section is devoted to the presentation and discussion of the numerical results. In this section an application to electrochemistry is presented, by studying the impact of correlation on chronoamperometric curves.

## 2-Results and discussion

### 2.1. Correlation-function-based theory

It has been demonstrated that for a 2D or 3D phase transformations, occurring by nucleation and growth, the fraction of the untransformed phase, $P(t)$, can be expressed in terms of either $f$-functions or correlation functions [30] according to

$$P(t) = 1 - \int_0^t I(t_1)dt_1 \int_{\Delta_{1t}} f_1(\boldsymbol{r}_1)d\boldsymbol{r}_1 + \int_0^t I(t_1)dt_1 \int_0^{t_1} I(t_2)dt_2 \int_{\Delta_{1t}} d\boldsymbol{r}_1 \int_{\Delta_{2t}} f_2(\boldsymbol{r}_1,\boldsymbol{r}_2)d\boldsymbol{r}_2 - \cdots$$

$$= 1 + \sum_{m=1}^{\infty} \frac{(-)^m}{m!} \int_0^t I(t_1)dt_1 \dots \int_0^t I(t_m)dt_m \int_{\Delta_{1t}} d\boldsymbol{r}_1 \dots \int_{\Delta_{mt}} f_m(\boldsymbol{r}_1, \dots, \boldsymbol{r}_m)\, d\boldsymbol{r}_m, \quad (1)$$

where $\Delta_{it} \equiv \Delta(t_i, t)$ ($i$=1,2,...,$m$) is the domain (a disk or a sphere for transitions in 2D or 3D space) transformed at time $t$ by a nucleus born at time $t_i$. The last expression in eqn.1 holds for $f_m$ functions symmetric with respect to the exchange of nucleus coordinates, $\boldsymbol{r}_i$. In eqn.1 $[I(t_1)dt_1 d\boldsymbol{r}_1][I(t_2)dt_2 d\boldsymbol{r}_2] \dots [I(t_m)dt_m d\boldsymbol{r}_m] f_m(\boldsymbol{r}_1, \dots, \boldsymbol{r}_m)$ is the probability of finding a nucleus born between times $t_i$ and $t_i + dt_i$ in the volume element $d\boldsymbol{r}_i$ at $\boldsymbol{r}_i$, irrespective of the position of the others $N - m$ nuclei, with $N$ total number of nuclei. For instance, $f_2(\boldsymbol{r}_1, \boldsymbol{r}_2)$ is the radial distribution function for the pair of nuclei (1,2). For a homogeneous and isotropic system $f_1(\boldsymbol{r}_1) = 1$ and $f_2(\boldsymbol{r}_1, \boldsymbol{r}_2) = f_2(|\boldsymbol{r}_1 - \boldsymbol{r}_2|)$. In eqn.1, $P(t)$ is the probability a generic point of the space is not transformed by the growing phase at running time $t$. This probability is linked to the fraction of the transformed phase, $X(t)$, by the relation $P(t) = 1 - X(t)$. We point out that the nucleation rate, $I(t)$ in eqn.1, depends upon the stochastic process under study. In particular, if the transformation is compliant with the KJMA model (random nucleation in the uncovered portion of space and non-parabolic-type growths) then $I(t) = I_0(t)$ is the nucleation rate throughout the entire space, that is comprehensive of phantoms, and $f_m = (f_1)^m = 1$. In this case eqn.1 gives the KJMA kinetics $P(t) = \exp[-X_e(t)]$, with the extended volume $X_e(t) = \int_0^t I_0(\tau)|\Delta(t,\tau)|\,d\tau$, where $|\Delta|$ is the measure of $\Delta$. On the other hand, for parabolic growth, owing to phantom overgrowth, it is profitable not to include phantoms in the formulation by expressing the kinetics in terms of *actual* nucleation. In this case in eqn.1 $I(t) = I_a(t)$, $I_a(t)$ being the actual nucleation rate, and the $f_m$'s are different from unity because the positions of actual nuclei are correlated. Actual and phantom-included nucleation rates are related through the equation $I_a(t) = P(t)I_0(t)$.

Eqn.1 can be rewritten in terms of correlation functions, $g_m$, which are related to the $f$-functions through the following cluster expansion:

$$f_1(\boldsymbol{r_1}) = g_1(\boldsymbol{r_1})$$

$$f_2(\boldsymbol{r_1}, \boldsymbol{r_2}) = g_1(\boldsymbol{r_1})g_1(\boldsymbol{r_2}) + g_2(\boldsymbol{r_1}, \boldsymbol{r_2}) \quad\quad\quad\quad\quad\quad (2)$$

$$f_3(\boldsymbol{r_1}, \boldsymbol{r_2}, \boldsymbol{r_3}) = g_1(\boldsymbol{r_1})g_1(\boldsymbol{r_2})g_1(\boldsymbol{r_3}) + g_1(\boldsymbol{r_3})g_2(\boldsymbol{r_1}, \boldsymbol{r_2}) + g_1(\boldsymbol{r_2})g_2(\boldsymbol{r_1}, \boldsymbol{r_3})$$
$$+ g_1(\boldsymbol{r_1})g_2(\boldsymbol{r_2}, \boldsymbol{r_3})$$

....

Inserting eqn.2 in eqn.1 and exploiting the symmetry of the $g_m$ under exchange of coordinates of the nuclei, the following expression is eventually obtained

$$P(t) = \exp\left[\sum_{m=1}^{\infty} \frac{(-)^m}{m!} \int_0^t I(t_1)dt_1 \dots \int_0^t I(t_m)dt_m \int_{\Delta_{1t}} d\boldsymbol{r_1} \dots \int_{\Delta_{mt}} g_m(\boldsymbol{r_1}, \dots, \boldsymbol{r_m}) \, d\boldsymbol{r_m}\right]. \quad (3)$$

This is the equation we employ for computing the kinetics of phase transformations with heterogeneous nucleation in the case of correlated nuclei. The computation is performed by considering terms up to second order in correlation functions. Owing to the quite low values of the nucleation densities (compared with typical liquid densities where the theory also applies) this approximation works well as shown through computer simulations [5,6]. Also, for a homogeneous and isotropic distribution of nuclei eqn.3 provides

$$P(t) \cong \exp\left[-\int_0^t I(t') \, dt' \int_{\Delta_{t't}} d\boldsymbol{r_1}\right.$$

$$\left. + \int_0^t I(t')dt' \int_0^{t'} I(t'')dt'' \int_{\Delta_{t't}} d\boldsymbol{r_1} \int_{\Delta_{t''t}} (f_2(|\boldsymbol{r_1} - \boldsymbol{r_2}|) - 1) \, d\boldsymbol{r_2}\right] \quad\quad (4a)$$

that is

$$P(t) \cong \exp\left[-\int_0^t I(t') |\Delta_{t't}|dt' + \int_0^t I(t')\,dt' \int_0^{t'} I(t'')dt'' \int_{\Delta_{t't}} d\boldsymbol{r_1} \int_{\Delta_{t''t}} f_2\left(|\boldsymbol{r_1}-\boldsymbol{r_2}|\right)d\boldsymbol{r_2}\right.$$

$$\left.-\int_0^t I(t')\,dt' \int_0^{t'} I(t'')|\Delta_{t't}||\Delta_{t''t}|\,dt''\right] \tag{4b}$$

where the cluster expansion eqn.2 was used and the time variables are indicated as $t'$ and $t''$. The probability $P(t)$ is computed by the requirement that in the region of space $\Delta_{t't}$ (centered at a generic point of space) no nucleation event takes place between $t'$ and $t' + dt'$ for $0 < t' < t$.

### 2.2. 3D-phase transformations with heterogeneous nucleation

In this section we deal with nucleation and growth of hemispherical nuclei on a solid surface. This process has been originally called 2D-1/2 growth mode, to stress that just 1/2 of the third dimension (positive z axis) is involved in the transformation. The growth law is given by the function $r_n(t, t')$, that is the radius of the nucleus at time $t$, with $t' < t$ being the birth time of the nucleus. In the following, we employ eqn.4b for determining the volume of hemispherical nuclei growing on the substrate surface in the case of correlated nucleation. An important field of application of the present modeling is the Electrochemistry where spatial correlation among nuclei is significant. It is found that in electrochemical deposition, under high overpotential conditions, an exclusion zone for nucleation develops around each growing nucleus. Experimental data on the nearest neighbor distribution (nnd) functions do show strong deviations from the Poisson distribution which are explained on the basis of the "exclusion zone" approach [35-38]. This model exploits the equivalence between the growth rate of hemispherical nucleus (radius $r_n$) and that due to planar diffusion across a disk encompassing the nucleus center with radius $r_d > r_n$. This is the disk where further nucleation is prevented, i.e. the exclusion zone for nucleation. Computer simulations and analytical models of nnd's in the case of exclusion zone for nucleation, are found in good agreement with experimental data [38].

From the above, it stems that a "2D-exclusion zone" for nucleation implies a hard-disk correlation among actual nuclei. In this case $f_2(|\boldsymbol{r_1}-\boldsymbol{r_2}|) = f_{1,2}(r) \cong H(r - r_{d;1,2})$ [1] where the subscript reminds us that the radius of the exclusion zone depends on the couple (1,2) through their birth times (see below). For diffusional growth $r_n(t, t') = \sqrt{\beta(t - t')}$ and $r_d(t, t') = \sqrt{\gamma(t - t')}$ where $\beta$ and $\gamma$ are constants. The ratio $\rho = \frac{\gamma}{\beta} \geq 1$ is a measure of the degree of correlation among nuclei. Labeling the populations of nuclei with nucleation times, the radial distribution function eventually becomes $f_2(r, t', t'') = H(r - \sqrt{\gamma(t' - t'')})$ where $t' > t''$ is assumed.

---

[1] $H(\cdot)$ is the Heaviside step function.

To compute the volume of the deposit we determine the probability, $P(t, h)$, that a generic point at height $h$ from the substrate, is not transformed by the new phase up to time $t$ (Fig.1). As anticipated, the probability given by eqn.4b fulfills the requirement that in the region of space $\Delta_{t't}$ (centered at a generic point of space) no nucleation event takes place between $t'$ and $t' + dt'$. In the problem under investigation, nucleation is on the surface while nucleus growth is also along the surface normal. We link $P(t, h)$ to eqn.4 by noting that the requirement that a point at height $h$ from the surface is untransformed at time $t$ implies that no nucleation event takes place in time interval $dt'$, with $0 < t' < t$, in the region $\Delta_{t't}$ of measure $\pi R(t, t')^2$ given by

$$|\Delta_{t',t}| = \pi R(t, t')^2 = \pi[r_n(t, t')^2 - h^2], \tag{5}$$

where $h \leq h_{max} = r_n(t, 0)$. A graphical representation of these quantities is depicted in Fig.1. The stochastic problem is therefore equivalent to a stochastic process of dots in 2D-space; space integrals in eqns.4 are performed in 2D. The equation for $P(t, h)$ becomes

$$
\begin{aligned}
P(t, h) = \exp\Bigg[ &-\int_0^{\bar{t}} I(t')\, \pi R(t, t')^2 dt' \\
&+ \int_0^{\bar{t}} I(t')dt' \int_0^{t'} I(t'')dt'' \int_{\Delta_{t't}} d\boldsymbol{r}_1 \int_{\Delta_{t''t}} H\left(r - r_d(t', t'')\right)d\boldsymbol{r}_2 \\
&- \int_0^{\bar{t}} I(t')\, dt' \int_0^{t'} I(t'')|\Delta_{t't}||\Delta_{t''t}|dt''\Bigg],
\end{aligned}
\tag{6}
$$

where $r = |\boldsymbol{r_1} - \boldsymbol{r_2}|$ and, to simplify the notation, the $h$ dependence has been omitted in the $R$ and $|\Delta|$ functions. In eqn.6 the extreme of integration, $\bar{t}(t)$, satisfies the equation $R(t, \bar{t}) = 0$. As far as the nucleation rate is concerned, for uncorrelated nucleation the first integral has to be only retained in eqn.6 leading to the formulation already developed in ref.[34]. In fact, for constant nucleation rate $I_0$ (comprehensive of phantoms), use of eqn.5 in eqn.6 implies

$$P(t,h) = \exp\left[-\pi I_0 \int_0^{\bar{t}} [\beta(t - t') - h^2]\, dt'\right] = \exp\left[-\frac{\pi I_0 \beta t^2}{2}\left(1 - \frac{h^2}{\beta t}\right)\right]. \qquad (7)$$

The volume of the deposit, per unit area, is eventually computed through integration of eqn.7 over $h$ up to the maximum height $h_{max} = \sqrt{\beta t}$ according to:

$$V(t) = \int_0^{h_{max}} \left(1 - P(t,h)\right) dh = \frac{\sqrt{\beta t}}{2}\int_0^1 \frac{1}{\sqrt{\eta}}\left(1 - e^{-S_{ex}(t)(1-\eta)^2}\right) d\eta, \qquad (8)$$

where $\eta = h^2/\beta t$ and $S_{ex}(t) = \frac{\pi I_0 \beta t^2}{2}$ is the extended surface. Since both $I_0$ and $\beta$ are constants, $\sqrt{S_{ex}(t)}$ is in fact a reduced time.

Here, we present the computation for the more general case of correlated nuclei by expressing the kinetics in terms of *actual* nucleation rate: in eqn.6, $I(t) = I_a(t)$. In the "exclusion zone" model, nucleation is prevented in a disk of radius $r_d(t, t')$ attached to each nucleus. For random nucleation with nucleation rate $I_0$ (on the entire surface) and growth law $r_d(t, t') = \sqrt{\gamma(t - t')}$, the fraction of the surface where actual nuclei do not form is given by the KJMA theory, $\theta = 1 - e^{-\frac{1}{2}\pi I_0 \gamma t^2}$ and the nucleation rate reads $I_a(t) = (1 - \theta)I_0 = I_0 e^{-\frac{1}{2}\pi I_0 \gamma t^2}$. Under these circumstances the distribution of actual nuclei is correlated according to the hard-disk model. In the following, we consider this expression for $I_a(t)$ which is a very good approximation due to the negligible contribution of phantom overgrowth [39,40].

For a better reading of the paper, in the following we only describe the main formulae required for computing the second order terms in eqn.6. The details of the mathematical computations are reported in the Appendix.

We focus our attention on space integrals in eqn.6 and employ polar coordinates for the integration over $\boldsymbol{r}_1$:

$$\Omega(t',t'') = -|\Delta_{t't}||\Delta_{t''t}| + 2\pi \int_0^{R(t,t')} r_1 dr_1 \int_{\Delta_{t''t}} f_2 \, d\boldsymbol{r}_2 \ , \qquad (9)$$

where $R(t,t') < R(t,t'')$ and the dependence on $h$ has been omitted. Since $f_2 = H\big(r - r_d(t',t'')\big)$ the integral over $\boldsymbol{r}_2$ is the portion of the area $|\Delta_{t''t}|$ "accessible" to this nucleus under correlation constraints. Fig.2 shows a pictorial view of the geometrical meaning of the integral over $\boldsymbol{r}_2$. In the figure, the dashed disk of radius $r_d$, centered at the position of the first nucleus, $\boldsymbol{r}_1$, is not accessible to the second nucleus (located at $\boldsymbol{r}_2$).

In dependence of the relative size of $R(t,t'')$ and $r_d(t',t'')$ the following cases have to be considered; they are,

1) $r_d(t',t'') \leq R(t,t'')$;

2) $R(t,t'') \leq r_d(t',t'') \leq R(t,t'') + R(t,t')$;

3) $r_d(t',t'') \geq R(t,t'') + R(t,t')$.

These instances are depicted in Fig.2 where positions of nuclei 1 and 2 (i.e. $\boldsymbol{r}_1$ and $\boldsymbol{r}_2$) are constrained within the circles of radius $R(t,t')$ and $R(t,t'') > R(t,t')$, respectively. In terms of the $\Omega(t',t'')$ function and using the dimensionless variables $\zeta' = t'/t$ and $\zeta'' = t''/t$, the second order contribution in eqn.6 becomes

$$\chi = I_0^2 t^2 \int_0^{1-\eta} e^{-at^2\zeta'^2} d\zeta' \int_0^{\zeta'} e^{-at^2\zeta''^2} \Omega(\zeta',\zeta'') d\zeta'', \qquad (10)$$

where $a = \frac{1}{2}\pi I_0 \gamma$ and $\frac{\bar{t}}{t} = 1 - \eta$. The three cases above reported lead to a partition of the integration domain of eqn.10 in dependence of reduced height, $\eta$, and correlation degree $\rho = \frac{\gamma}{\beta}$. The partition of the integration domain $0 \leq \zeta' \leq 1 - \eta$, $0 \leq \zeta'' \leq \zeta'$ is illustrated in Fig.3 for the cases 1)-3). For $\rho > 1$ this domain is partitioned by the two straight lines of equations $\zeta_1''(\zeta') = \frac{\rho\zeta' - (1-\eta)}{\rho - 1}$ and $\zeta_2''(\zeta') = \zeta'\left(\frac{1+\rho}{\rho-1}\right)^2 - \frac{4\rho(1-\eta)}{(\rho-1)^2}$ (the details of the computation are reported in the Appendix). It is possible to show that the integral $\Omega(\zeta', \zeta'')$ is made up of two contributions: the first one is given by the product between $R(t, t')^2$ and either $r_d(t', t'')^2$ or $R(t, t'')^2$; the second one is the integral of the function $\omega'(t, t', t'', r) = \pi r_d(t', t'')^2 - \omega[R(t, t''), r_d(t', t''); r]$, over the $r$ variable, being $\omega[R(t, t''), r_d(t', t''); r]$ the overlap area between two circles of radius $r_d(t', t'')$ and $R(t, t'')$, at relative distance $r$. For particular configurations of the nuclei, the $\omega'$ term vanishes and this happens in all cases 1)-3). A graphical representation of the overlap term is also highlighted in Fig.2.

Once defined the integration domains, the computation of the argument of the exponential in eqn.6, from now on denoted as $V_{ex}(\eta, S_{ex})$, can be done. This is equal to $V_{ex} = \sum_{i=0}^{3} \chi_i$, where $\chi_0$ is the first order term in eqn.6 and $\chi_i$ ($i$=1,2,3) the contributions given by eqn.10 for the cases 1), 2), and 3). Though these functions depend on $\rho$, we omit $\rho$ in the notation. It is profitable to normalize lengths to the maximum radius, $\sqrt{\beta t}$, and areas to the square of the maximum radius, $\beta t$. Specifically, we define the reduced radii $X_0(\zeta', \zeta'') = \sqrt{\rho(\zeta' - \zeta'')}$ (i.e. $\frac{r_d(t', t'')}{\sqrt{\beta t}}$), $X_1(\zeta', \eta) = \sqrt{1 - \zeta' - \eta}$ (i.e. $\frac{R(t, t')}{\sqrt{\beta t}}$) and $X_2(\zeta'', \eta) = \sqrt{1 - \zeta'' - \eta}$ (i.e. $\frac{R(t, t'')}{\sqrt{\beta t}}$). Similarly, for reduced areas $\Delta_0 = \rho(\zeta' - \zeta'')$, $\Delta_1 = 1 - \zeta' - \eta$, $\Delta_2 = 1 - \zeta'' - \eta$ and $\bar{\omega} = \frac{\omega'}{\beta t}$. For two overlapping disks of radius $Y_1$ and $Y_2$ placed at distance $x$, the area of the non-overlapping portion of disk 2 (with radius $Y_2$) reads

$$\bar{\omega}[Y_1, Y_2; x] =$$
$$Y_2^2\left[\pi - \cos^{-1}\frac{x^2 - (Y_1^2 - Y_2^2)}{2xY_2}\right] - Y_1^2\cos^{-1}\frac{x^2 + (Y_1^2 - Y_2^2)}{2xY_1} + \frac{1}{2}\sqrt{4x^2Y_2^2 - (x^2 - (Y_1^2 - Y_2^2))^2}.$$

Therefore, according to Fig.2 we get $\bar{\omega}_1 = \bar{\omega}[X_2, X_0; x]$ and $\bar{\omega}_2 = \bar{\omega}[X_0, X_2; x]$ for cases 1) and 2) respectively. The full expressions of the $\chi_i$'s computed in the Appendix are:

$$\chi_0(\eta, S_{ex}) = -\sqrt{\pi} S_{ex}(1-\eta) \frac{1}{\sqrt{\rho S_{ex}}} \mathrm{erf}\left((1-\eta)\sqrt{\rho S_{ex}}\right) + \frac{1}{\rho}\left(1 - e^{-\rho S_{ex}(1-\eta)^2}\right) \tag{11a}$$

$$\chi_1(\eta, S_{ex}) = 4 S_{ex}^2 \int_0^{\frac{1-\eta}{\rho}} e^{-\rho S_{ex}\zeta'^2} d\zeta' \int_0^{\zeta'} d\zeta'' e^{-\rho S_{ex}\zeta''^2} \left[-\Delta_1\Delta_0 + \frac{2}{\pi}\int_{y_1(\zeta',\zeta'',\eta)}^{X_1(\zeta',\eta)} \overline{\omega}_1 x_1 dx_1\right]$$

$$+ 4 S_{ex}^2 \int_{\frac{1-\eta}{\rho}}^{1-\eta} e^{-\rho S_{ex}\zeta'^2} d\zeta' \int_{\zeta_1'(\zeta',\eta)}^{\zeta'} d\zeta'' e^{-\rho S_{ex}\zeta''^2} \left[-\Delta_1\Delta_0 + \frac{2}{\pi}\int_{y_1(\zeta',\zeta'',\eta)}^{X_1(\zeta',\eta)} \overline{\omega}_1 x_1 dx_1\right] \tag{11b}$$

$$[\chi_2(\eta, S_{ex}) + \chi_3(\eta, S_{ex})] = -4 S_{ex}^2 \int_{\frac{1-\eta}{\rho}}^{1-\eta} e^{-\rho S_{ex}\zeta'^2} d\zeta' \int_0^{\zeta_1''(\zeta',\eta)} e^{-\rho S_{ex}\zeta''^2} \Delta_1\Delta_2 \; d\zeta''$$

$$+ \frac{8 S_{ex}^2}{\pi} \int_{\frac{1-\eta}{\rho}}^{1-\eta} e^{-\rho S_{ex}\zeta'^2} d\zeta' \int_{\zeta_0''(\zeta',\eta)}^{\zeta_1''(\zeta',\eta)} e^{-\rho S_{ex}\zeta''^2} d\zeta'' \int_{y_2(\zeta',\zeta'',\eta)}^{X_1(\zeta',\eta)} \overline{\omega}_2 x_1 dx_1 \tag{11c}$$

with the functions $\zeta_0''(\zeta',\eta) = \left[\zeta'\left(\frac{1+\rho}{\rho-1}\right)^2 - \frac{4\rho(1-\eta)}{(\rho-1)^2}\right] H\left(\zeta' - \frac{4\rho(1-\eta)}{(1+\rho)^2}\right)$, $\zeta_1''(\zeta',\eta) = \frac{\rho\zeta'-(1-\eta)}{\rho-1}$ $y_1(\zeta',\zeta'',\eta) = -y_2(\zeta',\zeta'',\eta) = \sqrt{1-\zeta''-\eta} - \sqrt{\rho(\zeta'-\zeta'')}$, and $S_{ex} \equiv S_{ex}(t) = \frac{\pi I_0 \beta t^2}{2}$. The volume of the deposit, per unit area, is eventually computed through integration over $\eta = h^2/\beta t$:

$$V(t) = \frac{\sqrt{\beta t}}{2} \int_0^1 \frac{1}{\sqrt{\eta}}\left(1 - e^{V_{ex}(\eta, S_{ex})}\right) d\eta. \tag{12}$$

### 2.3 Numerical results

#### 2.3.1 Kinetics of deposited volume and surface coverage

We performed numerical computations of eqns.11-12 for $\rho \geq 1$. As far as the nucleation process is concerned, the case $\rho = 1$ is of particular significance as it is compliant with the analytical model by KJMA for progressive nucleation. In the KJMA approach actual nuclei are spatially correlated since they only form in the untransformed portion of the surface. In the framework of the "exclusion zone" model for correlated nucleation this requires, in fact, a $\rho = 1$. Furthermore, the kinetics with correlated nuclei is expressed in terms of actual nucleation rate and this allows one to get rid of phantoms from the mathematical formulation. This approach is required, in general, when dealing with parabolic-type growths where the KJMA kinetics does not strictly hold because of the overgrowth of phantoms. Nevertheless, for parabolic growth the KJMA model provides a very good approximation because deviations from the exact kinetics are less than a few percent [39,40]. It follows that for $\rho = 1$ the time dependence of the deposited volume should be in accord with eqn.8. The case of correlated nucleation with $\rho = 1$ is therefore a benchmark for assessing the accuracy of the second order approximation here employed.

For $\rho = 1$ the expression of $V_{ex}(\eta, S_{ex})$ simplifies since $\chi_2 + \chi_3 = 0$ and the last double integral in $\chi_1$ vanishes (Fig.3). In this case $\overline{\omega}_1$ is the only overlap contribution which, in turn, has been computed to be negligible (see below). Under these circumstances, for $\rho = 1$ $V_{ex}$ takes the following closed form

$$V_{ex}(\eta, S_{ex}) = \frac{1}{2}\left(1 - e^{-2S_{ex}(1-\eta)^2}\right) + 2\sqrt{\pi S_{ex}^3} \int_0^{1-\eta} e^{-S_{ex}\zeta'^2}\zeta'^2 \, \mathrm{erf}\left(\zeta'\sqrt{S_{ex}}\right) d\zeta'$$

$$-(1-\eta)\sqrt{2\pi S_{ex}}\left[\mathrm{erf}\left((1-\eta)\sqrt{2S_{ex}}\right) - 2^{-\frac{1}{2}}e^{-S_{ex}(1-\eta)^2} \, \mathrm{erf}\left((1-\eta)\sqrt{S_{ex}}\right)\right]. \qquad (13)$$

The deposited volume is obtained through integration over the $\eta$ variable after inserting eqn.13 in eqn.12. The result is shown in Fig.4, as solid symbols, for the dimensionless quantity $W(S_{ex}) = \frac{V(t)}{\sqrt{\beta}t}$ as a function of extended surface $S_{ex}$. Solid line in Fig.4 is the $W(S_{ex})$ kinetics obtained by exploiting the Poisson statistics according to eqn.8. The agreement between the two approaches indicates that truncation of the series up to second order terms provides a good description of the phase transition. In the inset of the same figure the difference between the computation of $W(S_{ex})$ by including and excluding the $\overline{\omega}_1$-containing term, has also been reported. This difference is found to be negligible, a situation already encountered in estimating the contribution due to the overgrowth of phantoms [30,39,40]. Fig.4 also shows the behavior of the fraction of surface covered by islands, $S$, which provides information on the kinetics of the film closure. Surface coverage is computed from eqn.6 by setting $h = 0$, that is $S(S_{ex}) = 1 - e^{V_{ex}(S_{ex},0)}$.

Computations of normalized $V(t)$'s have been performed for $\rho > 1$ by using eqn.12 and the $V_{ex}$ expression given by eqns.11. The kinetics, computed by either including or neglecting $\bar{\omega}_1$ and $\bar{\omega}_2$-containing terms, are shown in Fig.5 for $\rho = 4, 20$ and $40$. Similarly to the case $\rho = 1$, overlap contributions are found to be small even for these values of $\rho$. For the kinetics of Fig.5 the relative difference between transformed volumes, with and without overlap terms, is about 3%. This relative difference is between the integrals of the normalized volumes over the extended surface.

The behavior of the surface coverage is also displayed in Fig.5. It stems that the effect of hard-disk correlation is to reduce the rate of both film deposition and surface coverage as a function of $S_{ex}$. Such a behavior is due to the competition between the two relevant processes which rule the transformation, namely the nucleation of actual nuclei and the impingement. Correlation among nuclei implies a reduction in number density of actual nuclei which, in turn, leads to a less effective impingement among them. Therefore, nucleation and impingement have opposite effects on the kinetics. In terms of square reduced-time ($S_{ex}$), the impact of nucleation on the kinetics is more important. It is worth to discuss this behavior in connection to results on 2D-phase transformations by simultaneous nucleation. In this case the scaled independent variable is given by either $\pi N (Gt)^2$ [2] or $\sqrt{N} Gt$ [3] where $G$ is the radial growth rate and $N$ the number density of nuclei, i.e. of actual nuclei since phantoms are absent in such a nucleation process. It is found that for hard-disk correlation and linear growth, the transformation rate increases with correlation degree [2,3]. This behavior seems to be opposite to the one obtained for $S$ by the present modeling with progressive nucleation and parabolic growth (Fig.5). As noted above, this is due to the nucleation process which implies a lower surface density of actual nuclei. At time $t$ this density is given by $N_a(t) = I_0 \int_0^t e^{-\rho S_{ex}(t')} dt'$ that is rewritten as $N_a(t) = \frac{I_0 t}{2} \sqrt{\frac{\pi}{\rho S_{ex}(t)}} \mathrm{erf}(\sqrt{\rho S_{ex}(t)})$. The reduced variable $\tilde{S}_{ex}(t) = \pi N_a r_n^2(t, 0)$ is therefore similar to that previously employed for site saturation [2,3] and, accordingly, the behavior of surface coverage for progressive nucleation should exhibit similar trend with correlation degree, when plotted as a function of $\tilde{S}_{ex}(t)$. Fig.6 shows that this is the case for the kinetics of Fig.5 since the rate of 2D transformation increases with $\rho$. In particular, the scaled variable has been related to $S_{ex}$ through the relationship $\tilde{S}_{ex} = \pi N_a \beta t = \sqrt{\frac{\pi S_{ex}}{\rho}} \mathrm{erf}(\sqrt{\rho S_{ex}})$.

### 2.3.2 Application to electrodeposition

As reported in the introduction, the present modeling is suitable for describing kinetics of electrodeposition at large overpotentials, where nuclei are correlated owing to the development of exclusion zones for nucleation [35, 38]. Recently, experimental data on the nearest neighbor distribution (nnd) function have been successfully interpreted considering hard-disk correlation among actual nuclei with $\rho > 1$ [38 and references therein]. In ref.[38] an analytical model has been presented for the pair-correlation function and the nnd function in case of progressive nucleation with correlation. Although the importance of spatial correlation has been recognized in electrochemical nucleation, a kinetic theory of electrodeposition with time dependent non-Poissonian nucleation has not been formulated so far.

By employing the present computation we are now in the position to model the typical quantity measured in electrodeposition experiments, namely the current density, $J$. This quantity is linked to the amount of deposited material per unit area, $V(t)$, by Faraday's law: $J(t) = zF \frac{1}{v_c} \frac{dV(t)}{dt}$ where $v_c$ is the specific volume of the deposited species, $z$ its valence and $F$ Faraday's constant. In terms of $S_{ex}$, the current density reads

$$J(S_{ex}) = A \left[ S_{ex}^{3/4} \frac{dW(S_{ex})}{dS_{ex}} + S_{ex}^{-1/4} \frac{W(S_{ex})}{4} \right], \tag{14}$$

where $W(S_{ex}) = \frac{V(t)}{\sqrt{\beta t}}$ is given by eqn.12 and $A = \frac{zF}{v_c} (8\pi I_0 \beta^3)^{1/4}$ is a constant. The current density for several values of correlation parameter, $\rho$, have been reported in Fig.7 in the normalized form $J//J_{max}$ vs $\tau/\tau_{max}$, where $\tau = \sqrt{S_{ex}}$ and the subscript *max* denotes values at maximum. This representation is commonly employed in the literature to plot experimental potentiostatic curves [41]. The current density exhibits a peak shape and a long time scaling according to $\tau^{-1/2}$. The results displayed in Fig.7 indicate that the higher the correlation among nuclei the more broaden is the peak of the current density; the current peak for uncorrelated nucleation ($\rho = 1$) is the more narrow. In the range of $\rho$'s here investigated, the value of $\tau_{max}$ ($J_{max}$) is found to increase (decrease) with correlation degree, as displayed in the inset of Fig.7. The surface coverage at maximum current is in the range 0.45-0.6 depending on $\rho$. This outcome is explained by considering that the current peak is linked to the time at which overlap among islands becomes operative [42]. At large $\rho$, due to the reduced rate of actual nucleation, the kinetics of surface coverage is delayed and, with it, the appearance of the current maximum. The fact that the maximum is related to impingement among nuclei, rather than to the nucleation process, can be understood by considering the following argument. For an ensemble of nearly isolated nuclei the volume can be approximated by the extended volume $V(t) \cong \frac{2\pi}{3} \beta^{3/2} \int_0^t I_a(t')(t-t')^{\frac{3}{2}} dt'$, that is $V(\tau) \propto \tau^{5/2} \int_0^1 e^{-\rho\tau^2\xi^2}(1-\xi)^{\frac{3}{2}} d\xi$. Although this function is in reasonable agreement with the initial trend of the numerical computation - even in the region of the maximum - it is not equal to the exact kinetics since it does not admit any flex point, i.e. any maximum in current density. This is in accord with the interpretation of the maximum as due to the onset of island overlap.

For the sake of completeness, let us compare these curves with those predicted by the Scharifker and Hills model for progressive nucleation and diffusional growth [37]. The current density reads $J = zF \frac{c\sqrt{D}}{\sqrt{\pi t}} (1 - e^{-\rho S_{ex}})$ where $c$ is the concentration of the solution and $D$ the diffusion coefficient of the species in solution. In terms of the $S_{ex}$ dimensionless variable, this equation becomes

$$J(S_{ex}) = A \frac{2}{3} \left[ S_{ex}^{-1/4} \frac{(1-e^{-\rho S_{ex}})}{\rho} \right], \tag{15}$$

where used was made of the relationships between $\beta$ and $\gamma$ parameters and $c$ and $D$ [37]. In eqn.15 the constant $A$ is the same as in eqn.14. Noteworthy, in electrochemical deposition, owing to correlation among diffusional field of nuclei, the growth law is not strictly parabolic [43]. This effect is expected to be more important in simultaneous than in progressive nucleation since in the latter case nuclei are more dispersed on the surface because of exclusion zones for nucleation. The model discussed in the previous section is limited to the impact of spatial correlation on the kinetics. On the other hand, in the phenomenological model developed in [37] both effects are taken into account. It is possible to show that eqn.15 implies the scaling $\frac{J(z)}{J_{max}}$, with $z = \frac{S_{ex}}{(S_{ex})_{max}}$, which is independent of $\rho$. In fact, $\frac{J(z)}{J_{max}} = \frac{1}{z^{1/4}} \frac{1-e^{-\alpha_0 z}}{1-e^{-\alpha_0}}$ where $\alpha_0 \cong 2.34$ is the solution of the equation $e^{-\alpha_0}(4\alpha_0 + 1) = 1$. Accordingly, when plotted in this normalized form the kinetics for different values of $\rho$ all collapse on the same curve. On the contrary, the present approach with correlated nucleation does not obey, in general, this scaling because $V_{ex}$ is not a function of $\rho S_{ex}$.

For example, it is possible to show that in the limiting case $\rho \gg 1$ (without overlap terms) does show that the functional form of $W$ given by eqn.12 does not fulfill such a scaling. However, the results of Fig.7 indicate that this behavior is recovered in the limit of large values of both $\rho$ and $\rho S_{ex}$. In this limit, neglecting overlap terms, $V_{ex}$ scales as $V_{ex} \equiv V_{ex}(\frac{S_{ex}}{\rho}, \eta)$. It is possible to show that a function of the form $f(S_{ex}, \rho) \equiv f(\rho^k S_{ex})$, with $k$ power exponent, satisfies the scaling above in term of the reduced variable $z = \frac{S_{ex}}{(S_{ex})_{max}} = \frac{\rho^k S_{ex}}{\rho^k (S_{ex})_{max}}$. The scaled function eqn.15 (dashed line in Fig.7) provides a sort of mean value of the curves obtained by the present modeling with correlated nuclei and parabolic growth.

**Conclusions**

In this paper we have studied the role of spatial correlation in 3D-phase transformations taking place by heterogeneous nucleation and diffusional growth. The proposed approach makes use of correlation function technique with truncation of the series expansion up to second order terms. This provides a good approximation of the kinetics as checked by applying the theory to Poissonian nucleation for which the exact solution has been computed, analytically. It is shown that even for non-Poissonian nucleation ($\rho > 1$) the solution can be approximated by analytical expressions owing to the negligible contribution of the integrals over overlap areas. The model allows one to determine the evolution of the amount of the deposited material and surface coverage of the substrate. The theory is suitable for studying heterogeneous nucleation at electrode surface where correlation effects are important. Computations of potentiostatic transients in electrodeposition show that the peak broaden with increasing correlation and at large $\rho$ the current densities exhibit a scaling behavior. Although the model has been applied here to hemispherical nuclei, anisotropic shape, such as ellipsoid, can also be studied [44].

**Appendix**

In this appendix we report the computation of the $P(t, h)$ probability retaining terms up to the second order in the $g_m$ series. In the computation, $f_2(r, t', t'') \cong H\big(r - r_d(t', t'')\big)$ where $0 < t'' < t' < t$. We recall that $r_d(t', t'') = \sqrt{\gamma(t' - t'')}$, $r_n(t - t') = \sqrt{\beta(t - t')}$, $|\Delta_{t't}| = \pi R(t - t')^2 = \pi[\beta(t - t') - h^2]$ and $\rho = \frac{\gamma}{\beta} \geq 1$. Employing dimensionless quantities $\zeta' = t'/t$, $\zeta'' = t''/t$, $x = r/\sqrt{\beta t}$, $at^2 = \frac{1}{2}I_0\pi\gamma t^2 = \rho S_{ex}$ and $\eta = h^2/\beta t$, the first order term in the argument of the exponential in eqn.6 reads:

$$\chi_0 = -\pi\beta I_0 t^2 \int_0^{1-\eta} e^{-at^2\zeta'^2}(1 - \zeta' - \eta)d\zeta'$$

$$= -2S_{ex} \int_0^{1-\eta} e^{-\rho S_{ex}\zeta'^2}(1 - \zeta' - \eta)d\zeta' . \qquad (A1)$$

*A1-Determination of the integration domains*

Concerning the second order term we focus our attention on space integrals and employ polar coordinates for the integration over $\boldsymbol{r}_1$ and $\boldsymbol{r}_2$:

$$\Omega = 2\pi \int_0^{R(t,t')} r_1 dr_1 \int_0^{2\pi} d\vartheta \int_0^{R(t,t'')} f_2 r_2 dr_2 - |\Delta_{t't}||\Delta_{t''t}|, \qquad (A2)$$

where $R(t, t') < R(t, t'')$ and the dependence on $h$ has been omitted. Since $f_2 = H\big(r - r_d(t', t'')\big)$ the integral over $\boldsymbol{r}_2$ is equal to $|\Delta_{t''t}|$ without overlap area (Fig.2). The computation of eqn.S2 requires considering the following three cases:

**case 1)** : $r_d(t', t'') \leq R(t, t'')$

In this case we define the distance $z_1$ through the relation $z_1 + r_d(t', t'') = R(t, t'')$ with the constraints $0 < z_1 < R(t, t')$, namely

$$0 < R(t,t'') - r_d(t',t'') < R(t,t')$$

or, in dimensionless form,

$$0 < \sqrt{1 - \zeta'' - \eta} - \sqrt{\rho(\zeta' - \zeta'')} < \sqrt{1 - \zeta' - \eta}. \tag{A3}$$

Eqn.A2 is recast according to [2]

$$\Omega_1 = 2\pi \int_0^{z_1} r_1 dr_1 \int_0^{2\pi} d\vartheta \int_0^{R(t,t'')} f_2 r_2 dr_2$$

$$+ 2\pi \int_{z_1}^{R(t,t')} r_1 dr_1 \int_0^{2\pi} d\vartheta \int_0^{R(t,t'')} f_2 r_2 dr_2 - |\Delta_{t't}||\Delta_{t''t}|$$

$$= 2\pi \int_0^{z_1} [|\Delta_{t''t}| - \pi r_d(t',t'')^2] r_1 dr_1 + 2\pi \int_{z_1}^{R(t,t')} (|\Delta_{t''t}| - \omega) r_1 dr_1 - |\Delta_{t't}||\Delta_{t''t}|$$

$$= 2\pi \int_0^{R(t,t')} |\Delta_{t''t}| r_1 dr_1 - 2\pi \int_0^{z_1} \pi r_d(t',t'')^2 r_1 dr_1 - 2\pi \int_{z_1}^{R(t,t')} \omega r_1 dr_1 - |\Delta_{t't}||\Delta_{t''t}|$$

$$= -2\pi \int_0^{R(t,t')} \pi r_d(t',t'')^2 r_1 dr_1 + 2\pi \int_{z_1}^{R(t,t')} [\pi r_d(t',t'')^2 - \omega] r_1 dr_1$$

$$= -\pi^2 r_d(t',t'')^2 R(t,t')^2 + 2\pi \int_{z_1}^{R(t,t')} \omega'(t,t',t'',r_1) r_1 dr_1 . \tag{A4}$$

In eqn.A4 $\omega \equiv \omega[R(t,t''), r_d(t',t''); r_1]$ is the overlap area of two disks or radius $R(t,t'')$ and $r_d(t',t'')$ located at distance $r_1$ and $\omega' = (\pi r_d(t',t'')^2 - \omega)$ (Fig.2 in the main text). Also, eqn.A3 gives a constraint on the $\zeta'$, $\zeta''$ times in dependence on $\rho$. The first inequality in eqn.A3 implies $(\rho > 1)$

$$\zeta'' > \frac{\rho\zeta' - (1-\eta)}{\rho - 1} = \zeta_1'' , \tag{A5}$$

---

[2] In eqn.A4 use is made of the identity $2\pi \int_0^{R(t,t')} |\Delta_{t''t}| r_1 dr_1 = |\Delta_{t't}||\Delta_{t''t}|$.

while the second inequality in eqn.A3 is always true (i.e. $z_1 < R(t, t')$). In eqn.A4 the range of variation of $\zeta''$ and $x_1 = \frac{r_1}{\sqrt{\beta t}}$ are, respectively

$$\zeta_1'' \leq \zeta'' \leq \zeta'$$

and

$$y_1(\zeta', \zeta'', \eta) < x_1 < \sqrt{1 - \zeta' - \eta} \,,$$

with $y_1(\zeta', \zeta'', \eta) = \sqrt{1 - \zeta'' - \eta} - \sqrt{\rho(\zeta' - \zeta'')}$. Since $\zeta'' > 0$ and $\zeta'$ spans the interval

$0 \to (1 - \eta)$, from eqn.A5 we obtain (Fig.3 in the main text):

i) for $\zeta' > \zeta_0'$ : $\qquad \zeta_1'' < \zeta'' < \zeta'$ $\hfill$ (A6)

ii) for $\zeta' < \zeta_0'$ : $\qquad 0 < \zeta'' < \zeta'$ $\hfill$ (A7)

with $\zeta_0' = \frac{(1 - \eta)}{\rho}$ and $\zeta_1'' = \frac{\rho \zeta' - (1 - \eta)}{\rho - 1} = \frac{\rho}{\rho - 1}(\zeta' - \zeta_0')$.

In conclusion, for case 1) eqns.A6, A7 provide

$$\zeta_1'' H((\zeta' - \zeta_0')) \leq \zeta'' \leq \zeta',$$ $\hfill$ (A8)

where $H(\cdot)$ is the Heaviside step function.

**case 2)** : $R(t, t'') \leq r_d(t', t'') \leq R(t, t'') + R(t, t')$

In this case the arrangement of the disks is depicted in Fig.2 of the main text. It follows that for

$0 < r_1 < r_d(t', t'') - R(t, t'') = z_2$ the integration domain $\Delta_{t'' t}$ is within the "correlation disk" and

$$\int_0^{2\pi} d\vartheta \int_0^{R(t,t'')} f_2 r_2 dr_2 = 0. \tag{A9}$$

For $z_2 = r_d(t',t'') - R(t,t'') < r_1 < R(t,t')$ the integral provides

$$\int_0^{2\pi} d\vartheta \int_0^{R(t,t'')} f_2 r_2 dr_2 = \omega'. \tag{A10}$$

Summing the two contributions we eventually obtain

$$\Omega_2 = 2\pi \int_{z_2}^{R(t,t')} r_1 \omega' dr_1 - |\Delta_{t't}||\Delta_{t''t}|, \tag{A11}$$

where $\omega' = \pi R(t,t'')^2 - \omega[R(t,t''), r_d(t',t''); r_1]$.

In terms of normalized quantities the inequalities $R(t,t'') < r_d(t',t'') < R(t,t'') + R(t,t')$ read:

$$\sqrt{1 - \zeta'' - \eta} < \sqrt{\rho(\zeta' - \zeta'')} < \sqrt{1 - \zeta' - \eta} + \sqrt{1 - \zeta'' - \eta}. \tag{A12}$$

The first inequality in eqn.A12 gives ($\rho > 1$)

$$\zeta'' < \frac{\rho\zeta' - (1-\eta)}{\rho - 1} = \zeta_1''$$

which requires $\zeta_1'' > 0$, that is $\zeta' > \frac{(1-\eta)}{\rho} = \zeta_0'$.

The second inequality, $\sqrt{\rho(\zeta' - \zeta'')} < \sqrt{1 - \zeta' - \eta} + \sqrt{1 - \zeta'' - \eta}$, leads to

$$\zeta'' > \zeta_2''$$

with

$$\zeta_2'' = \zeta' \left(\frac{1+\rho}{\rho - 1}\right)^2 - \frac{4\rho(1-\eta)}{(\rho - 1)^2}. \tag{A13}$$

The inequality eqn.A12 leads to the conditions

i) for $\zeta_2'' > 0:$   $\zeta_2'' \leq \zeta'' \leq \zeta_1''$         (A14)

ii) for $\zeta_2'' < 0:$   $0 \leq \zeta'' \leq \zeta_1''$         (A15)

In conclusion, for case 2) $\zeta' > \frac{(1-\eta)}{\rho}$ and the range of variation of $\zeta''$ and $x_1 = \frac{r_1}{\sqrt{\beta t}}$ are

$$\zeta_0'' \leq \zeta'' \leq \zeta_1'' \tag{A16}$$

and

$$y_2(\zeta', \zeta'', \eta) < x_1 < \sqrt{1 - \zeta' - \eta} \tag{A17}$$

with  $\zeta_0'' = \zeta_2'' H(\zeta_2'')$  and $y_2(\zeta', \zeta'', \eta) = \sqrt{\rho(\zeta' - \zeta'')} - \sqrt{1 - \zeta'' - \eta}$.

**case 3)** : $r_d(t', t'') \geq R(t, t'') + R(t, t')$

In this case

$$\Omega_3 = -|\Delta_{t't}||\Delta_{t''t}|. \tag{A18}$$

In non-dimensional form the inequality above reads

$$\sqrt{\rho(\zeta' - \zeta'')} > \sqrt{1 - \zeta' - \eta} + \sqrt{1 - \zeta'' - \eta} \ . \tag{A19}$$

Eqn.A19 provides an inequality similar to the one obtained for case 2). Since in this case the inequality is reversed we get

$$0 \leq \zeta'' \leq \zeta_0'', \tag{A20}$$

where $\zeta_0'' = \zeta_2'' H(\zeta_2'') = \zeta_2'' H(\zeta' - \zeta_1')$ with $\zeta_1' = \frac{4\rho(1-\eta)}{(1+\rho)^2}$. Since $\frac{4\rho(1-\eta)}{(1+\rho)^2} > \frac{(1-\eta)}{\rho}$, also for the present case we can consider for $\zeta'$ the same domain as in case 2): $\frac{(1-\eta)}{\rho} < \zeta' < 1 - \eta$.

## *A2-Full expressions of the integrals*

In this section we determine the time integrals over $\Omega_i$ ($i$=1,2,3) for the three cases above and for $I(\zeta') = I_0 e^{-at^2\zeta'^2}$ with $a = \frac{1}{2}I_0\pi\gamma$. In terms of dimensionless variables the general form of the integrals read

$$\chi_i = t^2 \int I(\zeta')d\zeta' \int I(\zeta'')\Omega_i d\zeta''$$

$$= \pi^2\beta^2 t^4 I_0^2 \int e^{-at^2\zeta'^2}d\zeta' \int \frac{\Omega_i}{\pi^2\beta^2 t^2} e^{-at^2\zeta''^2}d\zeta''$$

$$= 4S_{ex}^2 \int e^{-\rho S_{ex}\zeta'^2}d\zeta' \int e^{-\rho S_{ex}\zeta''^2}\bar{\Omega}_i d\zeta'' \, , \tag{A21}$$

where $S_{ex} = \frac{\pi I_0\beta t^2}{2}$, $at^2 = \rho S_{ex}$ and $\bar{\Omega}_i = \frac{\Omega_i}{\pi^2\beta^2 t^2}$. Specifying the integration domains in eqn.A21, using rescaled lengths to maximum nucleus size, $\sqrt{\beta}t$, and setting $\zeta_0'' = \zeta_2'' H(\zeta_2'')$ one obtains

$$\chi_1 = 4S_{ex}^2 \int_0^{\frac{1-\eta}{\rho}} e^{-\rho S_{ex}\zeta'^2}d\zeta' \int_0^{\zeta'} d\zeta'' e^{-\rho S_{ex}\zeta''^2}\left[-\Delta_1\Delta_0 + \frac{2}{\pi}\int_{y_1(\zeta',\zeta'',\eta)}^{X_1(\zeta',\eta)} \bar{\omega}_1 x_1 dx_1\right]$$

$$+ 4S_{ex}^2 \int_{\frac{1-\eta}{\rho}}^{1-\eta} e^{-\rho S_{ex}\zeta'^2}d\zeta' \int_{\zeta_1''}^{\zeta'} d\zeta'' e^{-\rho S_{ex}\zeta''^2}\left[-\Delta_1\Delta_0 + \frac{2}{\pi}\int_{y_1(\zeta',\zeta'',\eta)}^{X_1(\zeta',\eta)} \bar{\omega}_1 x_1 dx_1\right] \tag{A22}$$

$$\chi_2 = 4S_{ex}^2 \int_{\frac{1-\eta}{\rho}}^{1-\eta} e^{-\rho S_{ex}\zeta'^2}d\zeta' \int_{\zeta_0''}^{\zeta_1''} d\zeta'' e^{-\rho S_{ex}\zeta''^2}\left[-\Delta_1\Delta_2 + \frac{2}{\pi}\int_{y_2(\zeta',\zeta'',\eta)}^{X_1(\zeta',\eta)} \bar{\omega}_2 x_1 dx_1\right] \tag{A23}$$

$$\chi_3 = -4S_{ex}^2 \int_{\frac{1-\eta}{\rho}}^{1-\eta} e^{-\rho S_{ex} \zeta'^2} d\zeta' \int_0^{\zeta_0''} e^{-\rho S_{ex} \zeta''^2} \Delta_1 \Delta_2 \, d\zeta'' \tag{A24}$$

and

$$\chi_2 + \chi_3 = -4S_{ex}^2 \int_{\frac{1-\eta}{\rho}}^{1-\eta} e^{-\rho S_{ex} \zeta'^2} d\zeta' \int_0^{\zeta_1''} e^{-\rho S_{ex} \zeta''^2} \Delta_1 \Delta_2 \, d\zeta''$$

$$+ \frac{8S_{ex}^2}{\pi} \int_{\frac{1-\eta}{\rho}}^{1-\eta} e^{-\rho S_{ex} \zeta'^2} d\zeta' \int_{\zeta_0''}^{\zeta_1''} e^{-\rho S_{ex} \zeta''^2} d\zeta'' \int_{y_2(\zeta',\zeta'',\eta)}^{X_1(\zeta',\eta)} \bar{\omega}_2 x_1 dx_1 \, , \tag{A25}$$

where $y_1 = -y_2 = \sqrt{1-\zeta''-\eta} - \sqrt{\rho(\zeta'-\zeta'')}$ ; $\Delta_2 = 1-\zeta''-\eta$ ; $\Delta_1 = 1-\zeta'-\eta$ ; $\Delta_0 = \rho(\zeta'-\zeta'')$; $\zeta_1'' = \frac{\rho\zeta'-(1-\eta)}{\rho-1}$. The areas $\bar{\omega}_i = \frac{\omega_i'}{\beta t}$ are given in rescaled form by setting $X_0(\zeta',\zeta'') = \sqrt{\rho(\zeta'-\zeta'')}$ (i.e. $X_0 = \frac{r_d(t',t'')}{\sqrt{\beta t}}$ ), $X_1(\zeta',\eta) = \sqrt{1-\zeta'-\eta}$ ( $X_1 = \frac{R(t,t')}{\sqrt{\beta t}}$ ) and $X_2(\zeta'',\eta) = \sqrt{1-\zeta''-\eta}$ (i.e. $X_2 = \frac{R(t,t'')}{\sqrt{\beta t}}$ ). To simplify the notation in the $\chi_i$'s we omit the functional dependences with $\rho$, $S_{ex}$ and $\eta$. According to the definition above, for two overlapping disks of radius $R_1$ and $R_2$ placed at distance $r$, the non-overlapping portion (area) of disk 2, $\bar{\omega}[R_2, R_1; r]$, is equal to

$$\bar{\omega}[R_1, R_2; r] = R_2^2 \left[ \pi - \cos^{-1} \frac{r^2-(R_1^2-R_2^2)}{2rR_2} \right] - R_1^2 \cos^{-1} \frac{r^2+(R_1^2-R_2^2)}{2rR_1} + \frac{1}{2} [4r^2R_2^2 - (r^2 - (R_1^2 - R_2^2))^2]^{1/2} \, .$$

In eqns.A22-A25, $\bar{\omega}_1 \equiv \bar{\omega}[X_2, X_0; x_1]$ and $\bar{\omega}_2 \equiv \bar{\omega}[X_0, X_2; x_1]$.

The probability that the point at height $h$ from the surface is untransformed at time $t$, is equal to

$$P(\eta, t) = e^{V_{ex}(\eta, S_{ex}(t))} \, ,$$

where $V_{ex} = \chi_0 + \chi_1 + \chi_2 + \chi_3$. The volume of the film, per unit area, is eventually computed through integration over $h$ or $\eta = h^2/\beta t$ variables according to:

$$V(t) = \int_0^{h_{max}} \big(1 - P(t,h)\big)\, dh = \frac{\sqrt{\beta t}}{2} \int_0^1 \frac{1}{\sqrt{\eta}} \big(1 - e^{V_{ex}(\eta, S_{ex}(t))}\big)\, d\eta. \tag{A26}$$

**Figure Captions**

Fig.1 Schematic representation of the critical region for the generic point of the space, $Q$, located at height $h < h_{mx} = r_n(t, 0)$. The probability that the point $Q$ is not transformed by the new phase up to $t$, is linked to the area of the critical region through eqns.5,6. The colored disk is the critical region (or capture zone) for $Q$: nucleation events occurring within the disk in time interval $t'$ - $t' + dt'$ are capable of transforming $Q$ before time $t$. The hemisphere centered at $N$ represents a nucleus just at the border of the critical region. The dashed circle, in red, is the correlation disk for nuclei $N$ and $N'$ which depends on birth time of both nuclei.

Fig.2 Configurations of a couple of nuclei with hard-disk correlation for the three cases discussed in the text. Solid concentric circles are the integration domains over coordinates of nuclei 1 and 2. In particular, the radii of the circles are $R(t, t')$ for nucleus 1 and $R(t, t'')$ for nucleus 2. Positions of the nuclei are within these circles with the additional constraint dictated by correlation. The dashed circle is the correlation disk with radius $r_d(t', t'')$. In the panels the overlap areas entering the computation of the probability function have been also highlighted. For cases 1) and 2)

$z_1 = R(t, t'') - r_d(t', t'')$ and $z_2 = -z_1$, respectively (see the Appendix for details).

Fig.3 Graphical representation of the integration domains for the $(\zeta'', \zeta')$ time variables for the three cases discussed in the text. Numbered colored regions refer to the various cases at $\rho > 1$. The boundaries of these domains depend on $\zeta'$ and $\eta$ variables and on correlation degree, $\rho$, through the functions $\zeta_1''(\zeta') = \frac{\rho \zeta' - (1 - \eta)}{\rho - 1}$ and $\zeta_2''(\zeta') = \zeta' \left( \frac{1 + \rho}{\rho - 1} \right)^2 - \frac{4\rho(1 - \eta)}{(\rho - 1)^2}$. In the graph, $\zeta_0' = \frac{(1 - \eta)}{\rho}$ and $\zeta_1' = \frac{4\rho(1 - \eta)}{(1 + \rho)^2}$. In the Poisson limit, nucleation is uncorrelated in the untransformed phase and only case 1) is to be considered as $\rho \rightarrow 1$.

Fig.4. The volume of the deposit, $W(S_{ex})$, as computed using the present model of correlated nucleation, is reported for the case $\rho = 1$ (solid symbols). Solid line is the solution obtained by exploiting Poisson statistics which requires inclusion of phantoms (eqn.8). At $\rho = 1$ the two approaches describe the same phase transformation. The inset shows the negligible contribution of the integral term containing the overlap area ($\bar{\omega}_1$). The dashed line is the surface coverage which gives information on the kinetics of film closure.

Fig.5. Behavior of the volume of the deposit for different correlation degrees. In these panels curves b) and a) are, respectively, the numerical computations with and without overlap terms in eqns.11.

Curve c) is the fraction of substrate surface covered by the deposit. Dashed lines are the volume (blue line) and surface coverage (black line) for $\rho = 1$ (KJMA-like transformation).

Fig.6. Behavior of the fractional coverage for several values of the correlation parameter, $\rho$, as a function of the $\tilde{S}_{ex}$ variable. When plotted in this variable the kinetics with correlated nuclei is qualitatively similar to that obtained for simultaneous nucleation in refs.[2,3]. The values of $\rho$ are reported on each curve.

Fig.7 Modeling of chronoamperometric transients during electrochemical deposition. Current densities and reduced time are both normalized to the maximum values. The correlation degree, $\rho$, is reported in the figure and increases according to the arrow direction. The dashed line is the normalized kinetics of Scharifker and Hills's model. The inset shows the behavior of $\frac{\tau_{max}(\rho)}{\tau_{max}(1)}$ and $\frac{J_{max}(1)}{J_{max}(\rho)}$ with correlation parameter, $\rho$.

**Figures**

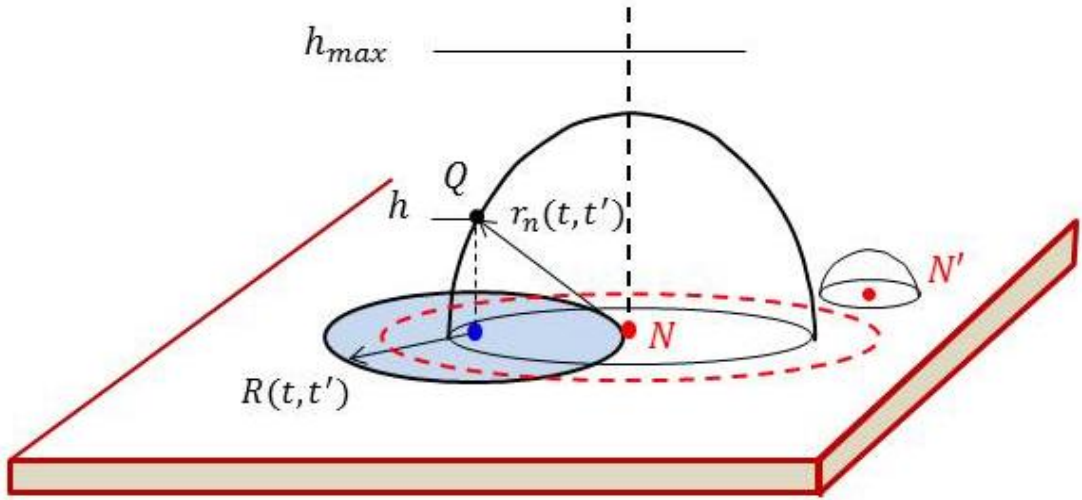

**Fig.1**

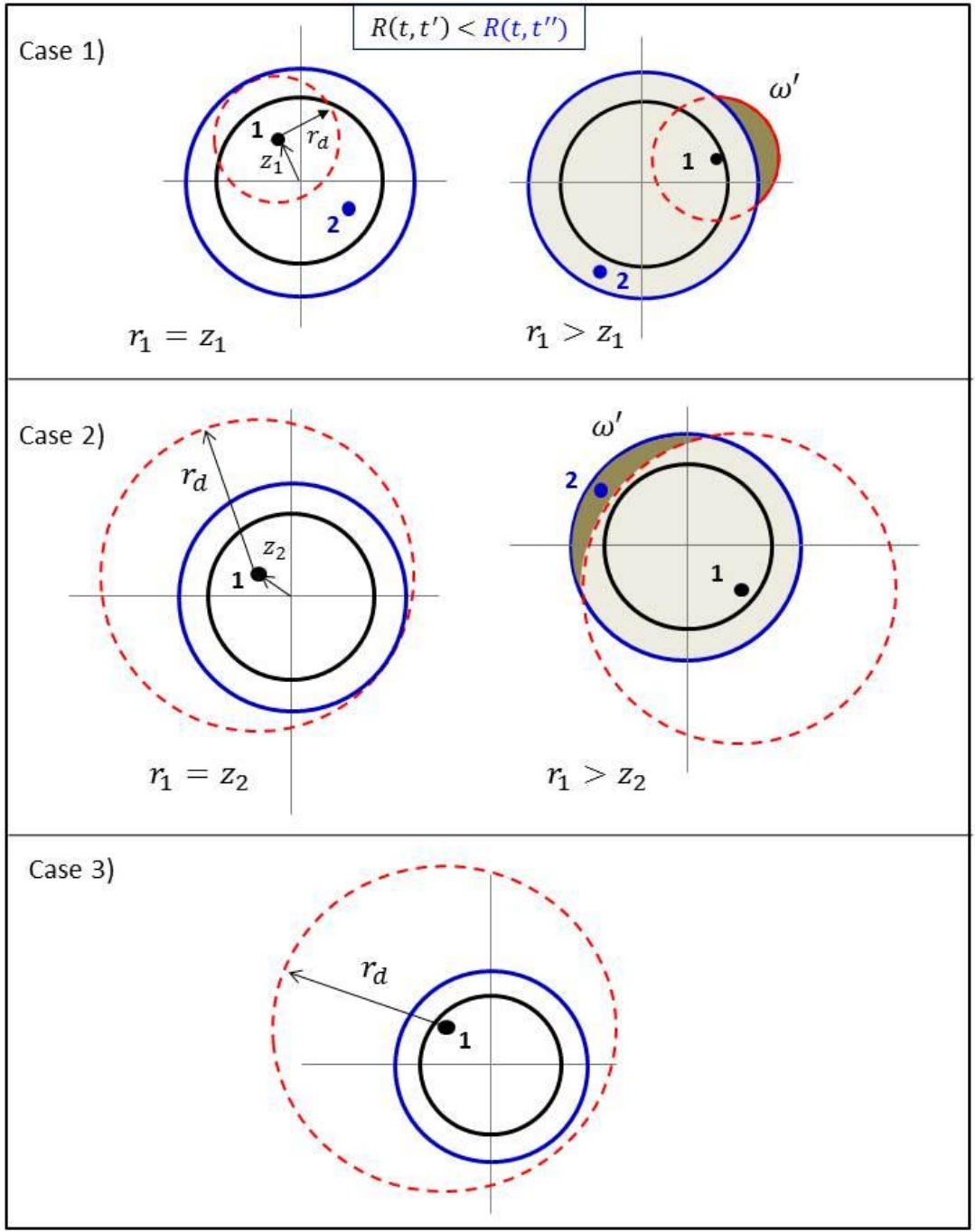

**Fig.2**

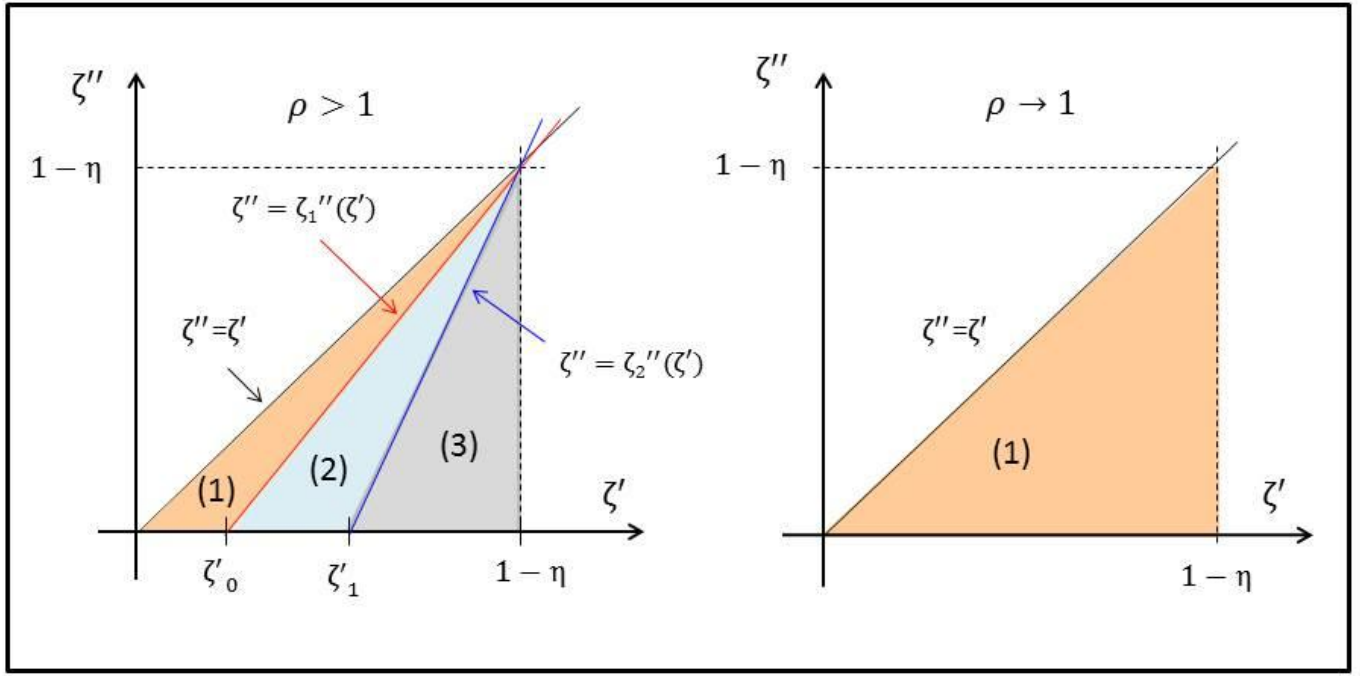

**Fig.3**

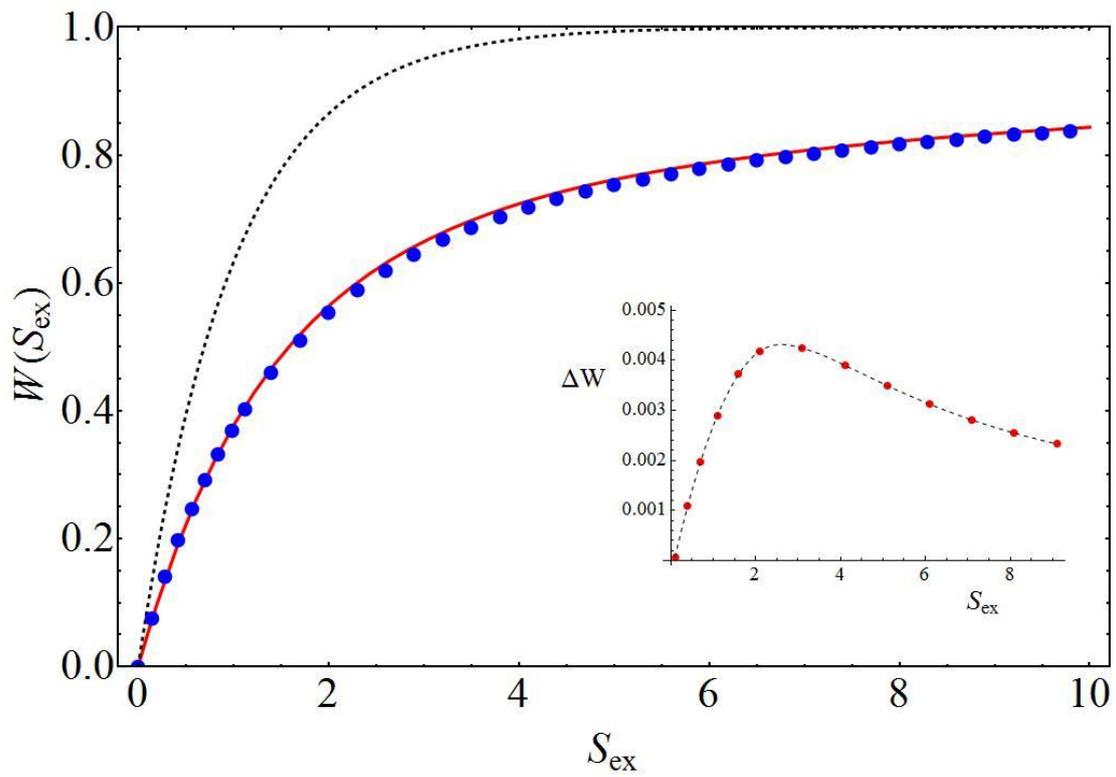

**Fig.4**

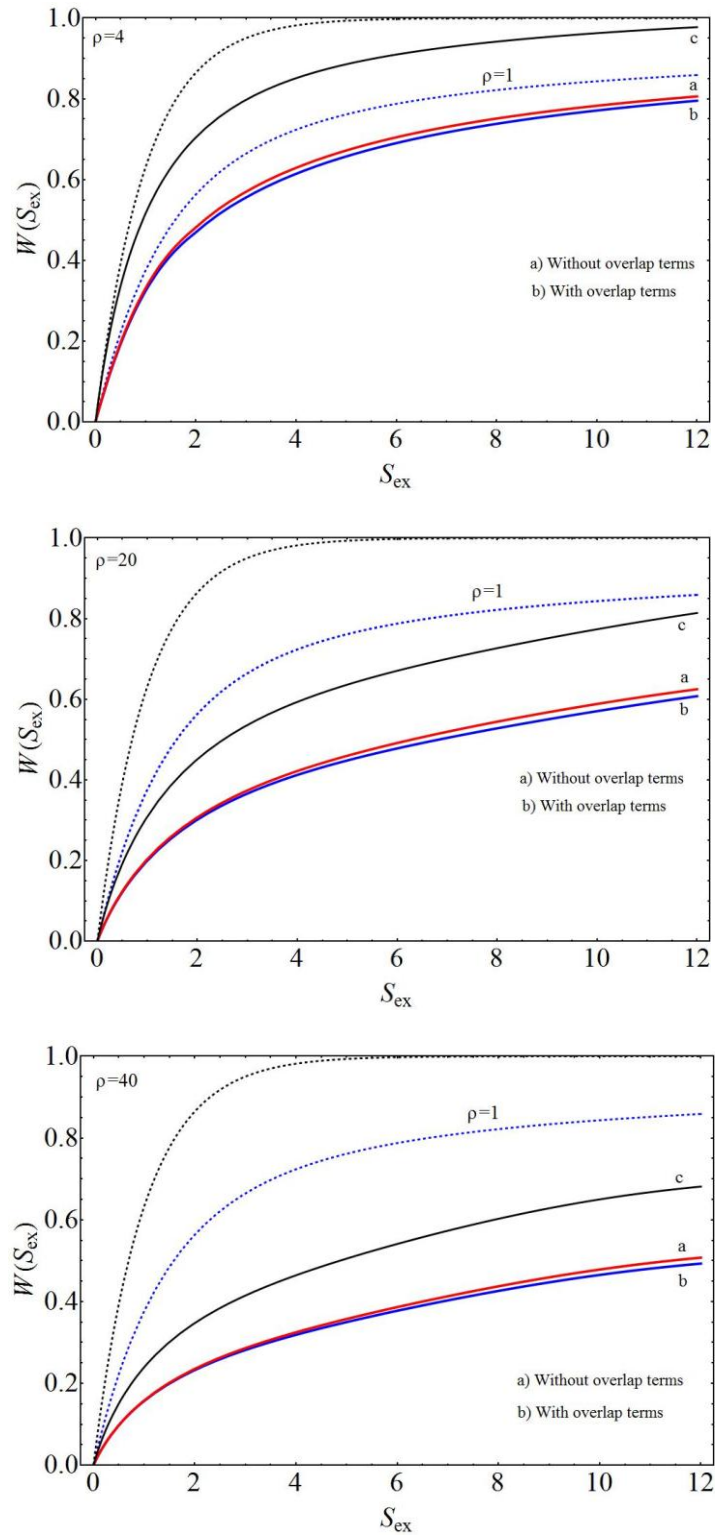

**Fig.5**

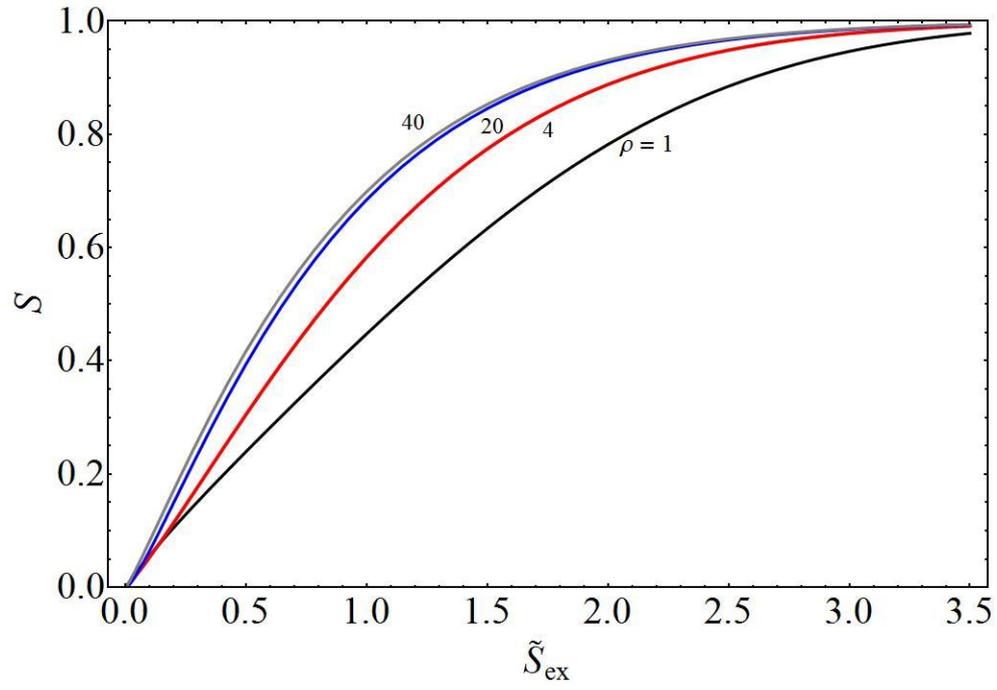

**Fig.6**

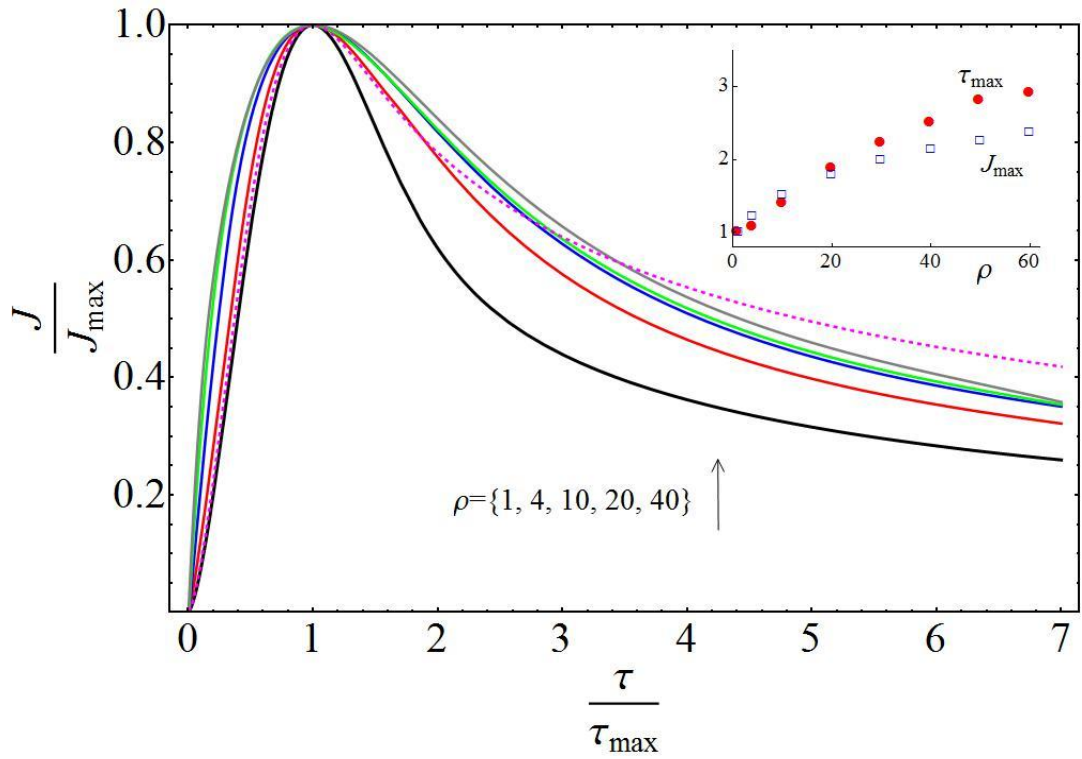

**Fig.7**